\documentclass[12pt,a4paper]{article}
\usepackage[utf8x]{inputenc}
\usepackage[T1,T2A]{fontenc}
\usepackage[russian,english]{babel}

\usepackage{amssymb,amsmath}
\usepackage{slashed}
\usepackage{mathtools}
\usepackage{feynmf}

\usepackage{mathrsfs}
\usepackage{graphicx}
\usepackage{simpler-wick}

\usepackage[export]{adjustbox}
\usepackage[colorinlistoftodos]{todonotes}
\usepackage[colorlinks=true, allcolors=blue]{hyperref}
\usepackage{hyperref}
\usepackage{authblk}

\usepackage{tikz}
\usepackage{subfigure}

\usepackage{indentfirst} 
\usepackage{cite}
\usepackage{gensymb}

\graphicspath{{Pics/}}

\def\cm{{\,\rm cm}}

\def\GeV{{\,\rm GeV}}

\title{The «Dark disk» model in the light of DAMPE experiment}
\author[1,*]{M.L. Solovyov}
\author[]{M.A. Rakhimova}
\author[]{K.M. Belotsky}
\affil[]{National Research Nuclear University MEPhI (Moscow Engineering Physics Institute), 115409, Kashirskoe shosse 31, Moscow, Russia}
\affil[*]{\small E-mail: max07s@mail.ru}
\date{}
\begin{document}
\maketitle
\begin{abstract}
    There are a lot of models considering the Dark Matter (DM) to be the origin of cosmic ray (CR) positron excess. However, they face an obstacle in the form of gamma-rays. Simple DM models tend to overproduce gamma-rays, leading to contradiction with isotropic gamma-ray background (IGRB). The <<dark disk>> model has been proposed to alleviate this contradiction. This work considers results of DAMPE experiment in the framework of the disk model. It is obtained that such a framework allows improving data fit considerably. \\[0.5cm]
\end{abstract}

\section{Introduction}

During the past decade, the anomalous behaviour of CR energy spectra was brought into the light. The positron excess, found by PAMELA \cite{Adriani:2008zr} and confirmed by AMS-02 \cite{PhysRevLett.110.141102,PhysRevLett.113.121101}, the <<wide>> and <<peak-like>> excesses in electron plus positron spectrum of recent DAMPE experiment \cite{Cao:2017rjr,Jin:2017qcv} are amongst the most well-known ones. The simple solutions seem to not work for these puzzles, as they remain unsolved. 


There are plenty of models considering DM of different nature, and there is a great freedom in defining its properties. Therefore, it is quite easy  to introduce a model with decaying or annihilating DM to account for the CR puzzles. And the possibility to probe for the new physics of the Dark Matter makes these models even more appealing.

The main alternative to DM models involves the pulsars as the cause of the excesses. Recent works in this field face the constraints \cite{2017Sci...358..911A,Shao-Qiang:2018zla} from gamma-radiation observed around the pulsars. Though attempts to solve the problems in this way continue (e.g., \cite{2020arXiv201002844L}).

However, DM models are also subject to the constraints. One of them is set by the gamma-ray data\cite{Belotsky:2016tja}. Photons are inevitably produced in the process of DM particle decay or annihilation via the final state radiation (FSR) process. And simple halo-distributed DM models dedicated to CR anomaly description tend to overproduce gammas, resulting in contradiction to Isotropic Gamma-ray Background (IGRB) data provided by Fermi-LAT\cite{Ackermann:2014usa}.

To resolve it, we develop the so-called <<dark disk>> model with unstable DM distributed in disk\cite{Belotsky:2018vyt}. This assumption helps to exclude gamma-rays from the outer regions of DM halo, that can not make a contribution to the observable charged particles fluxes.

In our previous work, we have found the IGRB data to constrain the halo models even in the case of broad electron plus positron excess in DAMPE data\cite{2019PDU....2600333B}. In this work we try to apply the <<dark disk>> model to this case. Sec.~\ref{InSet} provides the model description, Sec.~\ref{Res} contains the obtained results and everything is summed up in Sec.~\ref{Con}.

\section{Initial settings}
\label{InSet}

We consider DM particles with mass $m_X=1800$ GeV to be able to annihilate via 3 leptonic channels ($e^+e^-,\, \mu^+\mu^-,\, \tau^+\tau^-$) which branching ratios along with annihilation cross section $\langle\sigma v\rangle$ are the model parameters. For DM distribution, we use two density profiles:
\begin{itemize}
    \item Read's profile \cite{2008MNRAS.389.1041R}
    \begin{equation}
    \rho(r,z)=\rho_{0r}\text{exp}\left(-\cfrac{r}{R_R}\right)\text{exp}\left(-\cfrac{z}{zc}\right)
    \label{Read}
    \end{equation}
    \item NFW profile \cite{Navarro:1996gj} with cut-off along the $z$-axis
    \begin{equation}
     \rho(r,z)=\begin{cases}
    \cfrac{\rho_{0N}}{\frac{r}{R_s}\left(1+\frac{r}{R_s}\right)^2},& z\leq zc,\\
    0,& z>zc
    \end{cases}
    \label{NFW}
    \end{equation}
\end{itemize}
where $r$ and $z$ are coordinates in cylindrical coordinate system, $zc$ is the disk half-thickness, $\rho_{0r}=1.32 \GeV \cm^{-3}, \rho_{0N}=0.25 \GeV \cm^{-3}$, which corresponds to the local DM density of 
$0.4 \GeV \cm^{-3}$, $R_R=6.96\text{ kpc, } R_s=24$ kpc.

We use Pythia to calculate the initial spectra of electrons, positrons and gammas. The GALPROP code is used to propagate the first two of them and obtain their near-Earth spectra, as well as the secondary gamma flux. The prompt radiation flux is obtained by
\begin{multline}
    \Phi_{\text{prompt}}(E_\gamma)=\frac{dN_\gamma}{dE_\gamma}\frac{\langle\sigma v\rangle}{4}\times\\
    \times\frac{1}{\Delta\Omega}\int^{100\text{ kpc}}_0\int^{90^\circ}_{20^\circ}\int_0^{2\pi}\frac{1}{4\pi r^2}\left(\frac{\rho}{M_X}\right)^2r^2\cos(\theta)\, dr \, d\theta\,  d\phi\, ,
    \label{Fpromt}
\end{multline}
where $\frac{dN_\gamma}{dE_\gamma}$ is the gamma-ray spectrum per one act of annihilation, $M_X$ is the mass of DM particle, $\Delta\Omega$ is the solid angle $(l\in[0;2\pi]$, $b\in[20\degree;90\degree]$) corresponding to the region of the Fermi-LAT analysis. 

We use the total $e^+e^-$ background from \cite{Niu:2017lts}, which was obtained as the best-fit background model for a variety of cosmic-ray data.

To obtain the values of branching ratios and the process cross-section, we minimize the following expression for $\chi^2$:
\begin{gather}
    \chi^2=
    d^{-1}\left[\, \sum_{\substack{\rm DAMPE
    }}\frac{\left(\Delta \Phi_{e}\right)^2}{\sigma_e^2}+
    \sum_{\substack{\rm Fermi
    }} \frac{\left(\Delta\Phi_{\gamma}\right)^2}{\sigma_{\gamma}^2}\, H\left(\Delta\Phi_{\gamma}\right) \right].\label{chi2}
\end{gather}
Here $\Delta\Phi_i\equiv\Phi_{i}^{\rm (th)}-\Phi_{i}^{\rm (obs)}$, 
$\Phi_{i}$ are the predicted (\textit{th}) and measured (\textit{obs}) fluxes for $i = e,\gamma$ denoting $e^+e^-$ or gamma points respectively, $\sigma_{i}$ denotes the corresponding experimental errors and $d$ denotes the number of statistical degrees of freedom, which includes all the relevant DAMPE and Fermi-LAT data points. The first sum in Eq.~\eqref{chi2} goes over the DAMPE data points and the second sum goes over the Fermi-LAT data points. DAMPE points are taken in the range $20\div 1600$ GeV. Since we do not try to fit the gamma-ray data, but rather not to go over the experimental limits, the terms in the second sum are non-zero only when $\Phi_{\gamma}^{\rm (th)}>\Phi_{\gamma}^{\rm (obs)}$, which is ensured by the Heaviside step function $H$.

We use two different approaches for the minimization procedure. In first, called <<combined fit>>, we just simply minimize expression \ref{chi2}.  In the second, called <<$e$-fit>>, we minimize only the first sum in the expression \ref{chi2} and only after that, using the obtained parameters, we calculate total chi-square value.

\section{Results}
\label{Res}

\begin{figure}[h!] 
\centering
  \subfigure[]{%
    \includegraphics[width=.85\textwidth]{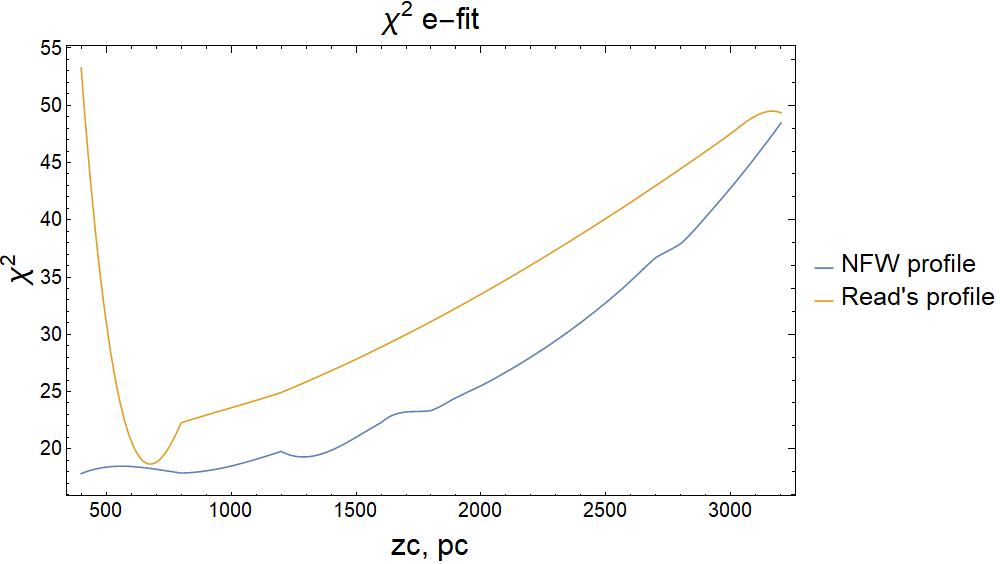} \label{fig:1} 
  } 
  \hfill
  \subfigure[]{%
    \includegraphics[width=.85\textwidth]{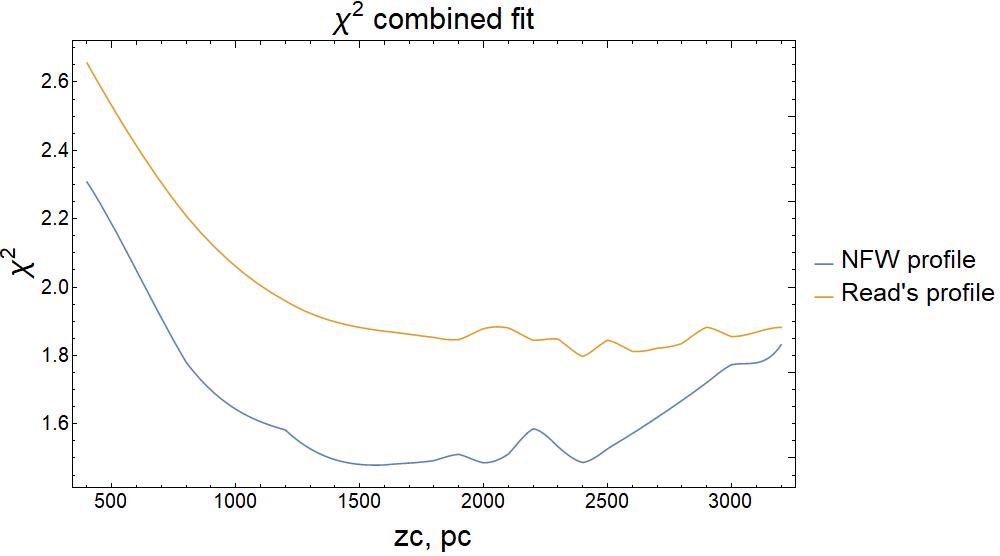} \label{fig:2} 
  } 
  \caption{Graphs for $\chi^2$ values in dependence of the disk half-width in case of e-fit~\subref{fig:1} and combined fit~\subref{fig:2}. Blue line is used for NFW density profile, the orange one -- for Read's density profile.}  \label{graphchi}
\end{figure}

Fig.~\ref{graphchi} illustrates the correlation between $\chi^2$ values and the disk half-width.

In the case of <<e-fit>> the best results are obtained with $zc\approx 750$ pc. However, one can clearly see that the quality of fit is still not satisfactory at all, although still better than one for the thick disks and halo. On the other hand, <<combined fit>> gives much better results with the minimum of $\chi^2$ of around 1.6 for the disk half-width in the range of $1500\div2000$ pc. However, in the case of AMS-02 positron fraction best fits were obtained with $zc=400$ pc. Unexpectedly, the NFW density profile with cut-off produces better results, than Read's profile, over the whole considered region. We suppose it to be due to higher production of low-energy electrons and positrons for NFW, which helps it to account for the lower energy region of the spectra. The line in the graphics breaks are mainly caused by the change of degree-of-freedom number (as we dynamically calculate it to include only those Fermi-LAT datapoints, where we have the excess) and interpolation errors.


\begin{table}[h!]
    \centering
    \scalebox{1.5}{%
   \begin{tabular}{|c|c|c|}
    \hline
    Fit | Model & Halo & Disk\\
    \hline 
    e-fit & 203 (0.53) & 17.85 (0.52) \\
    \hline
    combined fit & 3.8 (2.1) & 1.48 (1.20) \\
    \hline
\end{tabular}}
    \caption{The best-fit values of $\chi^2$ for different DM models and approaches for the minimization procedure. The values in brackets are obtained using only electron-positron part of Eq.~\eqref{chi2}.}
    \label{tab:comp}
\end{table}

Table \ref{tab:comp} contains the best-fit values of chi-square in contrast to the ones, obtained for the halo case. The comparison revealed that the dark disk model allows achieving the same accuracy in positron description, as the halo model, while giving less contradiction with IGRB. In both cases, combined fit improves the fit quality, but still not enough to overcome the discrepancy.

\section{Conclusion}
\label{Con}

We continue our research of DM explanation of the CR puzzles. In this work, we have applied the <<dark disk>> model to the case of the wide excess of positrons plus electrons in DAMPE data. We have obtained that it helps to lessen the contradiction with cosmic gamm-ray data. However, it is achieved at the cost of thicker disk, compared to the case of low energy positron anomaly of AMS-02.

In our future works we plan to run such analysis for the different masses of initial particle, try different reaction modes and to attempt to describe AMS-02 and DAMPE data simultaneously.

\section*{Acknowledgments}

The work was supported by the Ministry of Science and Higher Education of the Russian Federation
by project No 0723-2020-0040 ``Fundamental problems of cosmic rays and dark matter''.
Also we would like to thank R.Budaev, A.Kirillov and M.Laletin for their contribution at the early stage of this work.

\bibliographystyle{JHEP}
\bibliography{Bibliography}

\end{document}